\documentclass[prd,singlecolumn,nofootinbib,superscriptaddress,notitlepage]{revtex4-1}  
\usepackage{amsmath,amssymb,mathrsfs,bm,setspace,xspace,soul,empheq}         
\usepackage{graphicx}        
\usepackage[tight]{subfigure}       
\usepackage{color}   
 \usepackage{geometry}      
\usepackage[colorlinks=true]{hyperref}             
\hypersetup{ 
    bookmarks=true,         
    unicode=false,          
    pdftoolbar=true,        
    pdfmenubar=true,        
    pdffitwindow=false,     
    pdfstartview={FitH},    
    pdftitle={My title},    
    pdfauthor={Author},     
    pdfsubject={Subject},   
    pdfcreator={Creator},   
    pdfproducer={Producer}, 
    pdfkeywords={keyword1} {key2} {key3}, 
    pdfnewwindow=true,      
    colorlinks=true,       
    linkcolor=magenta, 
    citecolor=blue,        
    filecolor=magenta,      
    urlcolor=cyan           
} 

\newcommand{\al}{\alpha}

\newcommand{\de}{\epsilon} 
\newcommand{\e}{\epsilon}

\newcommand{\bra}[1]{\left\langle #1\right|}
\newcommand{\ket}[1]{\left| #1\right\rangle}

\newcommand{\beq}{\begin{equation}}
\newcommand{\eeq}{\end{equation}}
\newcommand{\ba}{\begin{array}{ccc}}
\newcommand{\ea}{\end{array}}

\def\bea{\begin{eqnarray}}
\def\eea{\end{eqnarray}}
\newcommand{\bml}{\begin{multline}}

\newcommand{\eeqm}{\end{multline}}
\newcommand{\bsp}{\begin{split}}
\newcommand{\esp}{\end{split}}

\renewcommand{\b}[1]{{\bf #1}}

\newcommand{\mc}{\mathcal}

\newcommand{\ts}{\thinspace{}}
\newcommand{\req}[1]{Eq.\thinspace(\ref{eq:#1})} 

\newcommand{\rfig}[1]{Fig.\ts\ref{fig:#1}}

\DeclareMathOperator{\tr}{tr}

\begin{document}  

\title{Entanglement susceptibilities and universal geometric entanglement entropy}

\author{William Witczak-Krempa}     
\affiliation{Universit\'e de Montr\'eal}     

\begin{abstract}
The entanglement entropy (EE) can measure the entanglement between a spatial subregion and its complement,
which provides key information about quantum states. 
Here, rather than focusing on specific regions, we study how the entanglement entropy changes with small deformations of the entangling surface. This leads to the notion of entanglement 
susceptibilities. These relate the variation of the EE to the geometric variation of the subregion. 
We determine the form of the leading entanglement susceptibilities for a large class of scale invariant states,
such as groundstates of conformal field theories, and systems with Lifshitz scaling, which includes fixed points governed by disorder. We then use the susceptibilities to derive the universal
contributions that arise due to non-smooth features in the entangling surface: corners in 2d, as well as cones and trihedral vertices in 3d. 
We finally discuss the generalization to R\'enyi entropies.     
\end{abstract}

 \date{\today}
\maketitle   
\tableofcontents   

\section{Introduction}
The entanglement entropy (EE) can quantify how much entanglement exists between two regions of a given system. 
As it is defined from the state without input about its origin, 
it is a versatile quantity that can 
be evaluated in or out of equilibrium, whether the state is pure or mixed. It also comes with the freedom of choosing the partition, which can be
in real space or momentum space, among other choices. As most physical systems involve spatially local interactions, partitions in real space 
play an important role in their study. However, even restricting oneself to real space partitions, a plethora of partition geometries exists. 
A particularly common and useful choice
for the partitioning surface, called entangling surface, is the $(d-1)$-dimensional sphere in $d$ spatial dimensions. Specifically, this corresponds to
an interval in $d=1$, a circle in $d=2$, and a 2-sphere in $d=3$. For instance, when we consider the groundstate of a conformal field theory (CFT) the 
EE associated with such a spherical partition allows one to extract the number that counts degrees of freedom, in the sense that it decreases monotonically 
under renormalization group flow \cite{Wilczek94,Myers11,CH12,Casini17,Nishioka18}. Another important class of partitions consists of non-smooth entangling regions.      
For example, in $d=2$ the subregion can have corners, while in $d=3$    
it can have vertices as would arise for a cube. Such
entangling surfaces are particularly important because they are the most natural to consider on the lattice. 
For groundstates of a large family of gapless Hamiltonians, including conformal field theories, 
corners \cite{CH_rev,Hirata07,Kallin14,Devakul14,BuenoPRL,Bueno15,Faulkner15,BW15,Whitsitt16,Helmes16,Igloi18}  
and trihedral vertices \cite{Igloi18,Singh14,Hayward17,Bednik18} give rise to logarithmically divergent   
contribution to the EE with a universal prefactor. By universal, we mean that is independent of the microscopic scales and can be      
determined by the field theory associated with the universality class of the critical point or phase.                 

In this work, we adopt an approach that treats a large class of entangling surfaces on an equal footing. The idea is to consider 
how the EE changes as a function of small but otherwise general deformations of the entangling surface, as shown in \rfig{deform}. 
This leads to the concept of \emph{entanglement susceptibilities}. The second order entanglement susceptibility was studied in the context
of relativistic quantum field theories,
with a focus on the groundstates of CFTs \cite{Nozaki13a,Nozaki13b,Bhattacharya15,Faulkner15}.\footnote{In those works, the 
second order entanglement susceptibility was called ``entanglement density'' but we shall see that ``susceptibility'' is a more
appropriate designation, as noted in \cite{Faulkner15}.} As we argue below,       
the entanglement susceptibilities apply in a much broader context, and can be used to obtain powerful results for the EE of scale invariant states. 
In Section~\ref{sec:singular}, we show that the variation of the EE associated with a deformation containing certain singularities (like corners in 2 spatial
dimensions) receives a logarithmically divergent correction with a universal prefactor. More precisely, 2d scale invariant states are found to have
the following variation of the EE as a result of introducing a corner:
\begin{align}
  \delta S(\mbox{corner}) &= - \frac{C}{12} \, (\Omega -\pi)^2\log(L/\de) + \dotsb  
\end{align}
while in 3d we have
\begin{align}
  \delta S(\mbox{cone}) &= -\frac{\pi^2 C}{256} (\Omega-\pi)^2 \left(\log(L/\de)\right)^2  +\dotsb \\
  \delta S(\mbox{trihedral}) &= C \left( \frac{9}{32\sqrt{3} } + \frac{\pi}{4} \right) \left(\frac{2\pi}{3}-\theta \right) \log(L/\de) + \dotsb
\end{align}
where the dots denote the other contributions to the EE, including the area law.
These results hold in the nearly flat limit, i.e.\ when the entangling surface is almost smooth. This means that the opening angle $\Omega$ of 
the 2d corner or 3d cone is near $\pi$ (\rfig{corner-cone}), or that each of the faces of the 3d trihedral vertex has an 
opening angle near $2\pi/3$ (\rfig{smooth-tri}). 
The constant $C$, which is non-negative, is a universal quantity in the sense that is independent of the cutoff $\epsilon$ (and any other length scale). 
For the groundstates of CFTs, the above results 
for the corners and cones were known \cite{BuenoPRL,Faulkner15,Klebanov12}, while the one for trihedral vertices is new. For CFTs, $C$
is in fact a local observable,     
\begin{align}
  C= \frac{2\pi^2}{d+2}\, C_T
\end{align} 
where $C_T$ determines the 2-point correlator of the stress or energy-momentum tensor
in the groundstate:
\begin{align}\label{CT} 
 \langle T_{\mu\nu}(\b x)\, T_{\lambda\rho}(0) \rangle = \frac{C_T}{|\b x|^{2(d+1)}}\, \mathcal I_{\mu\nu,\lambda\rho}(\b x)\,, 
\end{align}
where $\mathcal I_{\mu\nu,\lambda\rho}$ is a dimensionless tensor structure fixed by symmetry \cite{Osborn94}. $C_T$ is non-negative by unitarity, and can be obtained by measuring the frequency dependent shear viscosity $\eta(\omega)$.
In Table~\ref{tbl:C} we give the value of $C$ for the scale invariant groundstates 
of various quantum systems: Dirac fermion and Ising CFTs, a disordered fixed point of Ising spins, and the $z\!=\! 2$ Lifshitz boson with Lagrangian density
$\mc L=(\partial_t \phi)^2-\kappa^2 (\nabla^2\phi)^2$.  
Needless to say that it would be interesting to obtain more insights about the universal coefficient $C$ for non-conformal systems. 
\begin{table} 
  \centering   
  \begin{tabular}{c||c|c|c|c}       
    $\;\; d\;\;$ & Dirac fermion & quantum critical Ising & critical diluted quantum Ising  & $z\!=\!2$ Lifshitz boson \\
    \hline \hline
    $1$ & 1 \cite{Wilczek94}  & 1/2 \cite{Wilczek94} & $(\ln 2)/2$ \cite{Refael04} & $3/2$ \cite{ChenPRB17} \\ \hline 
    $2$ & 3/32 \cite{BuenoPRL,Elvang15}  & 0.044368 \cite{Faulkner15,El-Showk14} & $5\sqrt{3}/(4\pi^2)$ \cite{Igloi18} & $1/\pi^2$ \cite{Moore06} \\ \hline 
    $3$ & $4/(5\pi^2)$ \cite{Klebanov12} & unknown & unknown & unknown  \\ \hline 
  \end{tabular}  
  \caption{{\bf Entanglement susceptibility coefficient $C$}. We give the universal coefficient $C$ 
    appearing in $\chi^{(2)}$, \req{chi-plane} and (\ref{eq:chi2-1d}), for scale invariant groundstates
of various quantum systems in spatial dimension $1,2$ and 3. The first two columns are described by CFTs at low energy (modulo marginal corrections for the quantum critical Ising model in $3d$). 
Note that massless Dirac fermions in $2d\, (3d)$ have $2\, (4)$ components. 
Next, the critical diluted Ising model is governed by a disordered 
fixed point. Last is the quantum $z=2$ Lifshitz boson \cite{Moore06,ChenPRB17}. 
\label{tbl:C}}
\end{table}     

The paper is organized as follows. In Section~\ref{sec:susceptibilities} we define the notion of entanglement susceptibilities for general 
states. In Section~\ref{sec:singular}, we apply our results to obtain the universal part of the EE of singular surfaces (corners, cones, trihedrals) in the nearly flat limit. In Section~\ref{sec:Renyi} we generalize our findings to the R\'enyi entanglement entropies. We end with a conclusion and outlook.   

\section{Entanglement susceptibilities} \label{sec:susceptibilities} 
\subsection{Entanglement entropy of a deformed region}
\begin{figure}[t]   
  \centering
  \includegraphics[scale=.58]{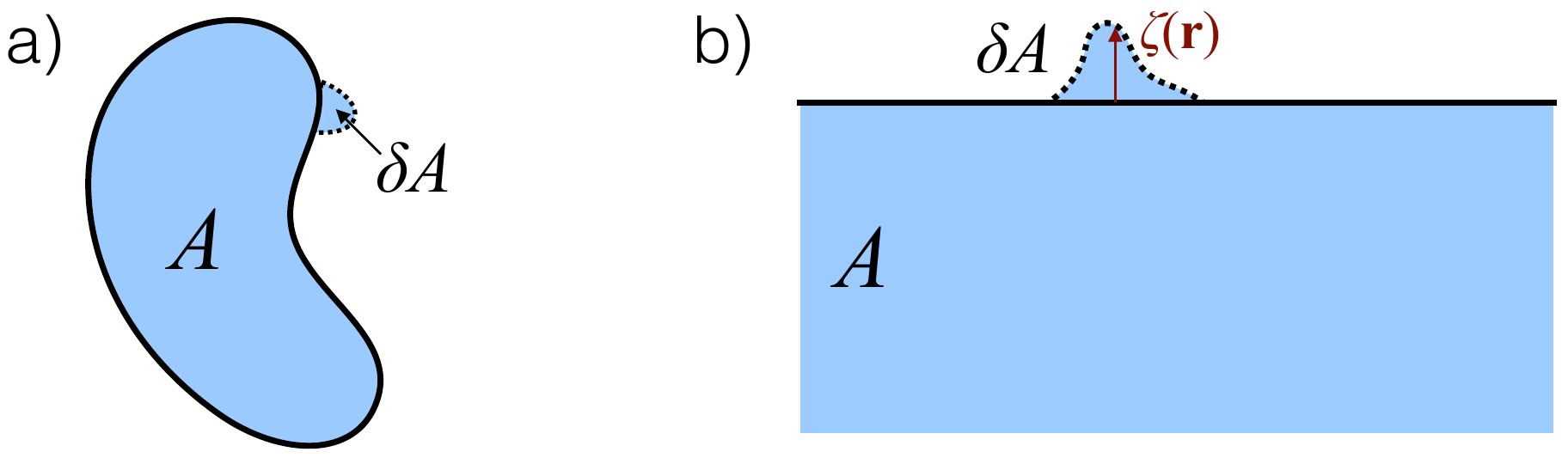}        
  \caption{\label{fig:deform} Deforming subregion $A$ to $A+\delta A$ in $d\!=\! 2$ spatial dimensions for $A$ a) compact, b) given by the half-plane.
The entanglement entropy of the deformed region can be expanded in powers of $\zeta(\b r)$, which defines the entanglement 
susceptibilities.}     
\end{figure}

Given a physical system described by a pure or mixed state defined by a density matrix $\rho$, we examine how the EE of a spatial subregion $A$ changes 
as we slightly deform it to $A+\delta A$, as shown in \rfig{deform}.  
In the case of disordered systems, it is often advantageous to average the EE over disorder
realizations, and our analysis applies in that case, with the replacement $S\to \langle S\rangle_{\rm dis}$.
The geometrical deformation $\delta A$ is defined by taking a point $\b r$ on the boundary of $A$, $\partial A$, and moving it to $\zeta(\b r) \b n(\b r)$,
where $\b n(\b r)$ is the unit normal at $\b r$ (\rfig{deform}b). 
We can then expand the EE of our state $\rho$, just as Faulkner \emph{et al} did for the groundstate of a CFT \cite{Faulkner15}:
\begin{align} \label{eq:S-exp}
  S(A+\delta A)= S(A) + \int_{\partial A}\!\! \mbox d^{d-1}\b r\,  \chi^{(1)}(\b r) \zeta(\b r)
  + \frac{1}{2!}\int_{\partial A}\!\! \mbox d^{d-1}\b r \int_{\partial A}\!\! \mbox d^{d-1}\b r'\, \chi^{(2)}(\b r,\b r') \zeta(\b r)\zeta(\b r') +\cdots 
\end{align}
which holds for $d\!>\!1$; a discussion of the $d\!=\! 1$ case is given in subsection~\ref{sec:intervals}.   
This expansion is akin to the expansion of the electric polarization in powers of the electric field, which defines the usual electric susceptibilities.
By analogy, we call the $\chi^{(\ell)}$ \emph{entanglement susceptibilities}. 
These not only depend on the state $\rho$, but also on the shape of region $A$. It thus seems like we can hardly conclude anything about the 
susceptibilities at this point.  
Interestingly, this is not the case: the 2nd order non-linear susceptibility must be non-positive,
\begin{align}  \label{eq:chi-neg}
  \chi^{(2)}(\b r,\b r')\leq 0
\end{align}
which follows from the strong subadditivity of entanglement entropy (SSA) \cite{Nozaki13b,Faulkner15}.  

To make further progress, we now assume that the state $\rho$ is translationnally invariant and lives in infinite space $\mathbb R^d$.
It now proves advantageous to pick a subregion $A$ that is left invariant under a maximal subgroup of translations, 
namely we take $\partial A$ to be entirely flat, i.e.\ a line in $d=2$ and a plane in $d=3$. Translation invariance then implies
that $\chi^{(1)}$ is constant and
\begin{align} 
  \chi^{(2)}(\b r,\b r')=\chi^{(2)}(\b r-\b r')
\end{align}
In addition, $\chi^{(\ell)}$ for $\ell$ odd must vanish if $\rho$ is pure. This is because the complement of $A+\delta A$ 
is $A-\delta A$, and vice versa, so purity implies $S(A+\delta A)=S(A-\delta A)$. This means that $S(A+\delta A)$ must be
even under $\zeta\to -\zeta$, ruling out all odd susceptibilities.    

In this work we are mainly interested in scale invariant states, such as groundstates of quantum critical Hamiltonians, for which an important simplification occurs. 
A scale transformation $\b r\to b\b r$ with $\zeta(\b r)\to b\zeta(b\b r)$ must leave $S(A+\delta A)$ invariant. This means that 
$\chi^{(\ell)}(\b r_1,\ldots,\b r_\ell)$ is homogeneous of order $-\ell d$. In particular, for $\ell=2$ this leads to one of the central
results of this work:
\begin{align} \label{eq:chi-plane}
  \chi^{(2)}(\b r-\b r') = - \frac{C}{|\b r-\b r'|^{2d}}
\end{align}
where $C\geq 0$ due to \req{chi-neg}. We have omitted terms that depend explicitly on
the cutoff as well as purely local contributions (contact terms) as they have no bearing on the universal properties.
The number $C$ appearing in the numerator of \req{chi-plane} is a universal measure of the long-range entanglement in the state 
$\rho=\ket{\psi}\! \bra{\psi}$. \req{chi-plane} was derived in the specific case of groundstates of CFTs \cite{Faulkner15}. 
In that context, $C$ is a local observable that characterizes the low energy excitations
of a Hamiltonian with groundstate $\ket\psi$. 
However, \req{chi-plane} holds
more broadly and provides a quantity, $C$, that can be compared among different theories as shown in Table~\ref{tbl:C}.   

It is difficult to make statements about the higher order susceptibilities. 
For instance, $\chi^{(4)}$ can in principle have an arbitrary dependence on the  
independent cross-ratios of four points.  

\subsection{Topological terms}
In our discussion of the entanglement susceptibilities, we have glossed over one subtlety, namely topological contributions.
Such terms do not contribute to the variation of the EE, $\delta S=S(A+\delta A)-S(A)$. Indeed, the deformation $\delta A$ is geometric in nature
as it does not alter the 
global properties of $A$ (or its boundary). However, such terms can be non-zero locally, i.e.\ for a portion of $\delta A$. For
example, let us examine the following contribution to the 2nd order variation of the EE for a scale invariant state when $A$ is a half-plane
in $d$ spatial dimensions:
\begin{align}
  \int_{\partial A}\!\! \mbox d^{d-1}\b r \int_{\partial A}\!\! \mbox d^{d-1}\b r'\, 
  \nabla_{\b r}\cdot\nabla_{\b r'}\!\left(\frac{ \zeta(\b r)\zeta(\b r')}{|\b r-\b r'|^{2(d-1)}}\right)
\end{align}
This integrand, which involves derivatives of $\zeta$, has the right scaling dimension and symmetry properties, just as \req{chi-plane}.
However, being a total derivative, it cannot contribute to $\delta S$. It is true that for generic deformations the integrand will be locally non-zero, but the entire integral nevertheless vanishes. 
One can consider other such terms, and their linear combinations, in order to understand what role they play for sub-features of $\delta A$. 
Since we are interested in the entire variation $\delta S$, we shall discuss them no further.  

\subsection{Spheres} 

We now turn to another geometry of importance namely when subregion $A$ is a ball of radius $R$ in $d$ dimensions, i.e.\ the entangling surface is the sphere $S^{d-1}$. Our starting point is again the expansion \req{S-exp}. Working with states that possess rotational and
translation symmetries, we conclude that the linear susceptibility is constant, $\chi^{(1)}(\b r)= f_0/R^d$, where $f_0$ is
a dimensionless constant. If we in addition require scale invariance, the second order entanglement susceptibility takes the form:
\begin{align} \label{eq:chi-sphere}
  \chi_{\rm sphere}^{(2)}(\b r -\b r') = - C\, \frac{f(|\b r -\b r'|/R)}{|\b r -\b r'|^{2d}} 
\end{align} 
where $f$ is a dimensionless function. As for the planar case, we have again omitted purely local contributions to the EE. 
For instance, one such local term would be $-\frac{C'}{R^{d+ 1}}\delta^{(d-1)}(\b r -\b r')$. 
This cutoff-independent term can be constructed for the sphere because we have a length scale, $R$,
to give it the correct units. 
Coming back to \req{chi-sphere}, $C$ is the same constant as the one that appears in the susceptibility of a planar entangling surface, \req{chi-plane}. 
Indeed, in the limit where $\b r\to \b r'$, the curvature becomes irrelevant and we must recover the answer of the half-plane:
\begin{align}
  f(u\ll 1)= 1+\dotsb
\end{align}
with the dots denoting terms that vanish in the small $u$ limit. 
Furthermore, $f(u)\geq 0$ for all $u$ by virtue of SSA, \req{chi-neg}. Unlike for the plane, we cannot set the function
$f$ to unity based on scale, rotational and translation invariance. However, $f\equiv 1$ does hold for the groundstate of a 
CFT \cite{Faulkner15}. This is because a half-plane region can be conformally mapped to a ball.
For such a groundstate, it was also shown that the universal part of the linear variation of the EE, $\delta S^{(1)}$,
vanishes \cite{Allais15}. This means that the universal part of the linear susceptibility $\chi^{(1)}$ vanishes. 

\subsection{Intervals in $d=1$} \label{sec:intervals} 
We now consider systems in $d=1$, and take subregion $A$ to be an interval of length $2R$. This in fact corresponds to a 0-sphere, denoted by $S^0$, of radius $R$.
The integrals over the entangling surface $\partial A$ in \req{S-exp} need to be replaced by discrete sums:
\begin{align} \label{eq:S-exp-1d}
  S(A+\delta A) = S(A) + \sum_{x\in\partial A} \chi^{(1)}(x) \zeta(x)+\frac{1}{2!}\sum_{x,x'\in\partial A}\chi^{(2)}(x,x')\zeta(x)\zeta(x')+ \dotsb
\end{align}
where the sums are over the end points of the interval, $x_1$ and $x_2$ with $x_2-x_1=2R$. As was the case above, SSA implies that
the 2nd order susceptibility is non-positive, \req{chi-neg}. If we restrict ourselves to scale and translation invariant states, 
we see that the first order susceptibility is constant: $\chi^{(1)}=f_0/R$. 
Further, $\chi^{(2)}$ is given by the discretized version of \req{chi-sphere}:  
\begin{align}
  \chi_{\rm interval}^{(2)}(x_i -x_j) = - \mc C \, \frac{1-\delta_{ij}}{(x_i -x_j )^2} 
  -\frac{\mc C'}{R^{2}}\delta_{ij} 
\end{align}
where $i,j\in\{1,2\}$ label the endpoints of the interval. 
We notice that the non-local term has the same form as the purely local one since $(1-\delta_{ij})/(x_i-x_j)^2=(1-\delta_{ij})/(2R)^2$,
so that $\chi^{(2)}_{\rm interval}(x_i-x_j)=-\tfrac{1}{4R^2}\left[\mc C+(4\mc C'-\mc C)\delta_{ij} \right]$. 

We can now determine the constants $\mc C, \mc C',f_0$ for translation and scale invariant states that have a logarithmically divergent EE: 
\begin{align} \label{eq:1d-EE}
  S(A)= \frac{C}{3}\ln\left(\frac{2R}{\de}\right) + b'
\end{align}
where $C\geq 0$, $\de$ is a short distance cutoff and $b'$ is a constant. For groundstates of CFTs, 
the constant $C$ is the Virasoro central charge $c$ \cite{Wilczek94},
but the scaling \req{1d-EE} also appears for groundstates of non-conformal theories, such as the Motzkin spin chain \cite{Bravyi12} or
the $z=2$ Lifshitz boson \cite{ChenPRB17,ChenJPA17}.   
Let us consider a small deformation of the interval $[x_1,x_2]$:      
\begin{align}
  x_1\to x_1 -\zeta_1\,, \qquad x_2\to x_2 + \zeta_2
\end{align}
where we have defined $\zeta_i\equiv \zeta(x_i)$. We can then expand the EE by replacing $2R\to 2R+\zeta_1+\zeta_2$ in \req{1d-EE}: 
\begin{align} \label{eq:S-1d-exp-direct}
  S(A+\delta A) = S(A)+ \frac{C}{3} \left(\frac{\zeta_1+\zeta_2}{2R}\right) - \frac{C}{6} \left(\frac{\zeta_1+\zeta_2}{2R}\right)^2 +\dotsb
\end{align}
Alternatively, we can compute the variation of the EE using \req{S-exp-1d}:
\begin{align} 
  S(A+\delta A) = S(A) +  \frac{f_0}{R}(\zeta_1+\zeta_2)-\frac{1}{8R^2}\left(2\mc C\zeta_1\zeta_2+4\mc C'(\zeta_1^2+\zeta_2^2) \right) +\cdots 
\end{align}
For this expansion to agree with \req{S-1d-exp-direct}, we need to set $f_0=\mc C/2=2\mc C'= C/6$, so that
\begin{align} \label{eq:chi2-1d}
  \chi^{(2)}_{\rm interval}(2R) = - \frac{C}{12R^2}
\end{align}
which is what was obtained for CFTs \cite{Nozaki13a} (up to a numerical prefactor due to a different convention), 
where $C$ is the central charge in that context.
We note that $\chi^{(2)}_{\rm interval}$ is negative as required by SSA.  



\section{Universal entanglement entropy of singular surfaces} 
\label{sec:singular}  

In this section, we use the entanglement susceptibility $\chi^{(2)}$ to obtain the EE of non-smooth regions. Important examples are corners in $d=2$,
cones and trihedral vertices in $d=3$. We shall focus on pure states $\rho=\ket{\psi}\bra{\psi}$ that possess scale, rotational and 
translation invariance. This includes, but is not restricted to, groundstates of CFTs. 
The general idea is to start with subregion $A$ being a flat plane, and then add a small non-smooth deformation $\delta A$.
The EE will vary as follows:
\begin{align}
  S(A+\delta A) &= S(A) + \delta S^{(2)} + \dotsb \\
  \delta S^{(2)} &= \frac{1}{2!}\int_{\partial A}\!\! \mbox d^{d-1}\b r \int_{\partial A}\!\! \mbox d^{d-1}\b r'\, 
\frac{-C}{|\b r-\b r'|^{2d}} \, \zeta(\b r)\zeta(\b r') \label{eq:main-int}
\end{align}
where the first variation vanishes as discussed above, and we have substituted the second-order entanglement susceptibility for a planar 
subregion $A$, \req{chi-plane}. Using this geometric expression we shall identify various universal coefficients in the EE. 
\begin{figure}[t]   
  \centering  
  \includegraphics[scale=.36]{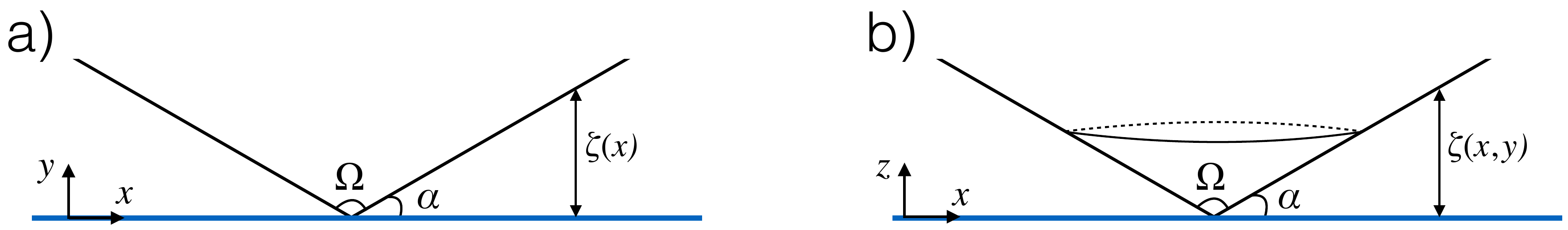}        
  \caption{\label{fig:corner-cone} a) Perturbing a linear entangling surface with a corner in 2d. 
b) Perturbing a half-planar entangling surface with a cone in 3d (side view).} 
\end{figure} 

\subsection{Corners in $d=2$}  
The corner deformation of the flat entangling region, shown in \rfig{corner-cone}a,   
is defined by 
\begin{align}
  \zeta(x) = |x|\tan\al 
\end{align}
where the angle $\alpha$ is taken to be small. The opening angle of the corner, whose tip lies at the origin, is $\Omega=\pi-2\alpha\simeq \pi$. 
The variation of the EE then reads:
\begin{align}
  \delta S^{(2)} = -\frac{C}{2} \times (2I_{\rm same}+2I_{\rm diff})
\end{align}
where we have decomposed the integral in \req{main-int} based on whether $x_1$ and $x_2$ map to the same 
edge, or to different ones.  
We first evaluate $I_{\rm same}$ on the $x,y>0$ edge of the corner:
\begin{align}
  I_{\rm same} = \int_0^\infty dx_1 \int_0^\infty dx_2 \frac{\zeta(x_1)\zeta(x_2)}{(x_1-x_2)^4}
\end{align}
We perform the $x_2$ integral first by splitting it into two parts in order to avoid
the divergence at coincidental points, $x_2=x_1$: $\int_0^{x_1-\e}dx_2 + \int_{x_1+\e}^\infty dx_2$. 
Here, $\e>0$ is the short distance cutoff. 
The result is 
\begin{align}
  I_{\rm same} = \tan^2\alpha \int_0^\infty dx_1 \frac{1+4(x_1/\e)^3}{6 x_1} = \frac{\al^2}{6}\log(L/\de) + \dotsb
\end{align}
We have performed the final integral with small and large distance cutoffs. 
We have also expanded at small $\de/L$, and replaced $\tan\al$ by $\al$ since we work in the nearly flat limit. 
The dots denote terms with integer powers of $\de/L$.
We thus see that we have recovered a logarithmically divergent contribution, with a cutoff-independent prefactor. 
We now calculate the contribution from opposite edges:
\begin{align}
  I_{\rm diff} &= \int_{-\infty}^0 \! dx_1\int_0^\infty\! dx_2\, \frac{\zeta(x_1)\zeta(x_2)}{(x_1-x_2)^4} 
= \int_{-\infty}^0\! dx_1\, \zeta(x_1)\, \frac{\tan\alpha }{6 x_1^2} \\ 
  &= \frac{\al^2}{6}\log(L/\de) + \dotsb
\end{align}
which is exactly the same as $I_{\rm same}$. Our final result for the variation of the EE is thus
\begin{align} \label{eq:sigma_corner}
  \delta S^{(2)} = - \frac{C}{12} \, (\Omega -\pi)^2\log(L/\de) + \dotsb 
\end{align}
which is expressed in terms of the corner's opening angle, $\Omega$. The dots denote purely non-universal terms.  
The prefactor of the logarithm is independent of any cutoff, including the short distance one. It is the product
of a theory dependent constant $C$ times a purely geometric $(\Omega-\pi)^2$ dependence. For the wide class of scale invariant
states considered here, corners thus contribute a logarithmically divergent contribution with a universal coefficient.
For CFTs, $C=\pi^2 C_T/2$ and 
\req{sigma_corner} agrees exactly with the conjecture of \cite{BuenoPRL,Bueno15}, proved by \cite{Faulkner15} using the entanglement
susceptibility but with a different deformation $\zeta(x)$ and method of regularization.    

\subsection{Cones in $d=3$}    
We now move to 3 spatial dimensions and consider one of the simplest singular entangling surfaces: 
a cone.
We work in the nearly flat limit; the corresponding deformation of the plane
is shown in \rfig{corner-cone}b. In cylindrical coordinates $(\rho, \phi,z)$, this reads
\begin{align}
  \zeta(\b r) = \rho \tan\al
\end{align} 
where $\alpha\ll 1$ so that the cone's opening angle $\Omega=\pi-2\alpha$ is near $\pi$.
The integral for $\delta S^{(2)}$ is
\begin{align}
  \delta S^{(2)} &= -\frac{C}{2} \int d\b r_1\int d\b r_2 \frac{\zeta(r_1)\zeta(r_2)}{[\rho_1^2 + \rho_2^2 -2\rho_1 \rho_2\cos(\phi_1-\phi_2)]^3} \\
    &= -\frac{C}{2}\,(2\pi)\int_0^\infty \rho_1 d\rho_1\int_0^\infty \rho_2 d\rho_2 \, \zeta(\rho_1)\zeta(\rho_2)\times 
(2\pi)\frac{\rho_1^4+4\rho_1^2 \rho_2^2 +\rho_2^4}{|\rho_1^2-\rho_2^2|^5} 
\end{align}
where in the second equality we have set $\phi_2=0$ by virtue of the rotation symmetry about the $z$ axis, 
and added an overall factor of $2\pi$ to
account for the $\phi_2$ integral. The $\phi_1$ integral was then performed exactly. The $\rho_2$ integral can then be performed 
by using the splitting method described above for the corner. Finally, we perform the $\rho_1$ integral over $[\de,L]$,
and expand at small $\de/L$ to get
\begin{align} \label{eq:cone-ee}
  \delta S^{(2)} 
= -\frac{\pi^2 C}{256} (\Omega-\pi)^2 \left(\log(L/\de)\right)^2  +\dotsb 
\end{align}
where the dots represent terms with integer powers of $\de/L$ (with single $\log(L/\de)$ prefactors for some of the terms). 
In contrast to corners in $d=2$, we find a log-squared $(\log(L/\de))^2$ singularity. For groundstates of CFTs,
a conical singularity of arbitrary opening angle contributes $-C_T h(\Omega)(\log(L/\de))^2$ to the EE, where the function $h$ is the same for \emph{all}
CFTs \cite{Klebanov12}. Expanding $h(\Omega)$ near $\pi$ yields \req{cone-ee} when we use $C=2\pi^2 C_T/5$. 

\subsection{Trihedral vertices in $d=3$}
\begin{figure}[t]   
  \centering
  \includegraphics[scale=.5]{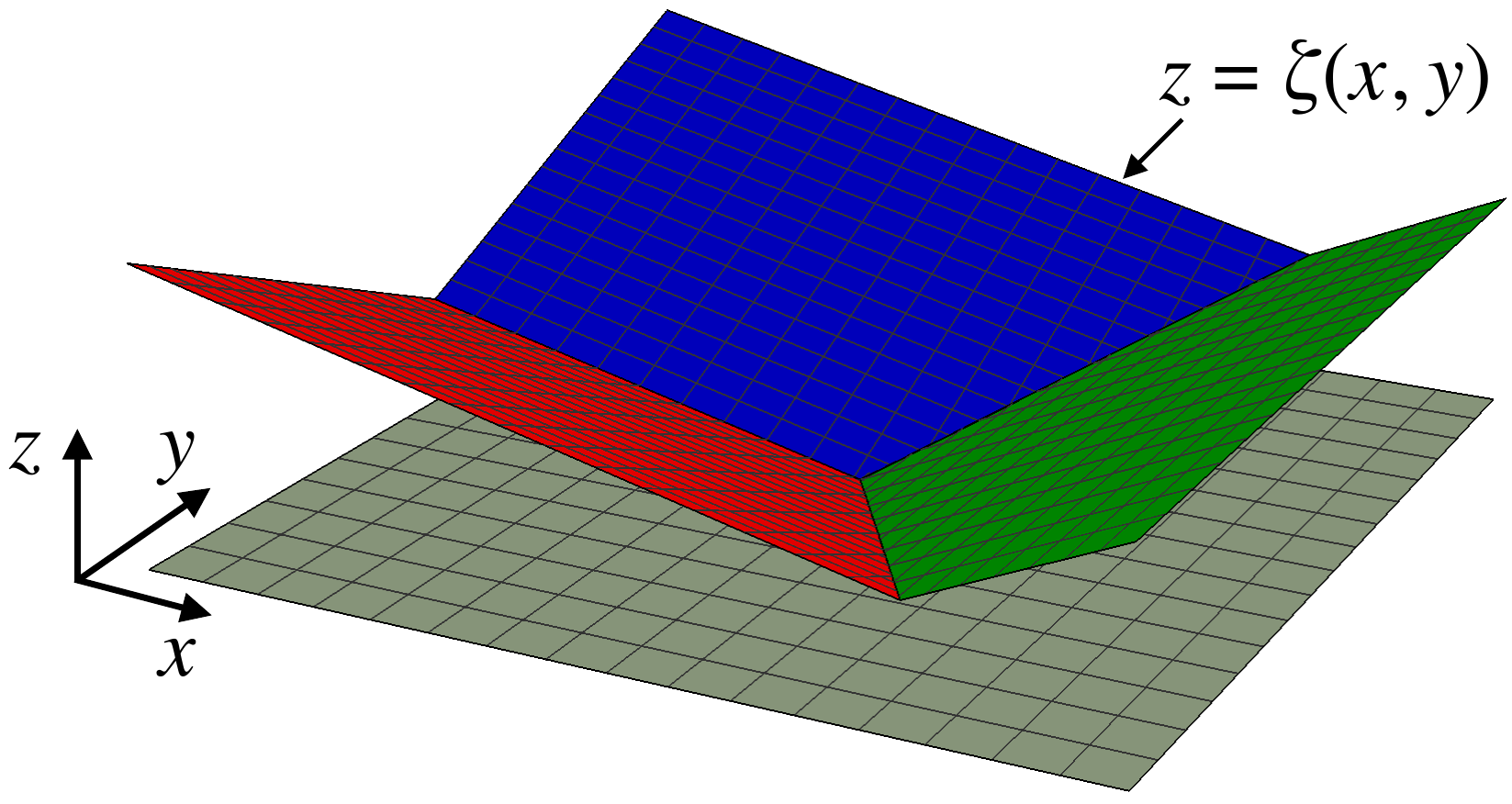}          
  \caption{\label{fig:smooth-tri} Perturbing a flat entangling  surface at $z=0$ with a nearly flat trihedral vertex, 
which is threefold rotation symmetric. 
The origin is at the the tip of the vertex. The 3 edges make an angle $\alpha$ with the $z=0$ plane.}
\end{figure}  
We now turn to our main calculation: the nearly flat trihedral in 3 spatial dimensions.
We proceed as for the corner by deforming a planar entangling surface into one containing a trihedral vertex, 
as shown in \rfig{smooth-tri}.  
We consider the symmetric nearly-flat limit, in which all 3 triangular faces have the same angle $\theta$
that approaches $2\pi/3$. 
Due to the piecewise nature of the entangling surface, the variation $\delta S$ can be decomposed into two contributions: 
$\b r_1,\b r_2$ are such that $\zeta(\b r_1)$ and $\zeta(\b r_2)$ lie either on the same triangular face, 
or on 2 neighboring ones: 
\begin{align} \label{eq:diff-same-decomp}
  \delta S^{(2)}= - \frac{C}{2}(3I_{\rm same} + 6I_{\rm diff})
\end{align}
where the integer prefactors come from the 3 faces, and 6 different ways of assigning $\b r_1$ and $\b r_2$ to neighboring faces, respectively. 
We first evaluate the integral for the first contribution:  
\begin{align} \label{eq:I_tri_same}
  I_{\rm same}= \int_0^\infty\!dx_1\int_{-\sqrt 3 x_1}^{\sqrt 3 x_1}dy_1 \int_0^{\infty} dx_2 \int_{-\sqrt 3 x_2}^{\sqrt 3 x_2}dy_2
  \frac{\zeta(\b r_1) \zeta(\b r_2)}{[(x_1-x_2)^2 + (y_1-y_2)^2]^3}   
\end{align}
where both $\zeta(\b r_1)$ and $\zeta(\b r_2)$ lie on the green face in \rfig{smooth-tri}: 
\begin{align}
  \zeta(\b r)= x\tan\al \, 
\end{align} 
and  $x>0$, $|y|<x\tan(\pi/3)$. 
Here, $\al$ is the angle between any of the 3 edges and the $z=0$ plane. The relation between $\al$ and the face's angle $\theta$ is
$\cos\al=\tfrac{1}{\sqrt 3} \tan(\theta/2)$. In the nearly flat limit, we thus have $\al^2\simeq 4(\frac{2\pi}{3}-\theta)/\sqrt 3$,
which vanishes as $\theta\to 2\pi/3$.     
We note that the $\tan^2\al$ angular dependence factorizes outside of the integral \req{I_tri_same}.
We first perform the $y_{1,2}$ integrals exactly:
\begin{multline} \label{eq:Isame_1}
  I_{\rm same}= \tan^2\al \!\int_0^{\infty}\!\! dx_1 \int_0^{\infty}\!\! dx_2\, x_1 x_2 \\
\times \frac{6 \sqrt{3} \left(x_1^2+x_2 x_1+x_2^2\right) \left[x_2 \left(\pi 
   \text{sgn}\left(x_{-}\right)-2 \tan^{-1}\left(\frac{x_{-}}{\sqrt{3}
   x_{+}}\right)\right)+2 x_1 \tan ^{-1}\left(\frac{\sqrt{3}
   x_{+}}{x_{-}}\right)\right]
+3 x_1 x_2 x_{-} +4 \sqrt{3} \pi  
   \left(x_2^3-x_1^3\right)}{16 x_{-}^5 (x_1^2+x_2 x_1+x_2^2)}   
\end{multline}
where $x_\pm=x_1\pm x_2$. 
We then perform the $x_2$ integral. To avoid the singularity at coincidental points, we split the integral
into two parts as above: $\int_0^\infty dx_2 = \int_0^{x_1-\de}dx_2 + \int_{x_1 + \de}^\infty dx_2$. The result can be obtained in closed form,
but we refrain from writing it due to its length. The final integral requires both short and long distance cutoffs but 
can also be obtained analytically. Expanding $I_{\rm same}$ at small $\e/L$, we get
\begin{align}
  I_{\rm same} = - \left(\frac{1}{16} + \frac{11\pi}{72\sqrt 3} \right) \alpha^2 \log (L/\de) +\dotsb \label{eq:I_same_fin}
\end{align}
where the dots represent terms with integer powers of $L/\de$. We have replaced $\tan\al\to\al$, which applies in the smooth limit. 
We thus see that a logarithmic term appears. 

We next evaluate the integral for $\b r_1$ and $\b r_2$ lying on neighboring faces. We choose $\b r_1$ such that 
$\zeta(\b r_1)$ is on the $y_1>0$ face 
(in blue in \rfig{smooth-tri}), while $\zeta(\b r_2)$ is on the $y_2<0$ one (in red in \rfig{smooth-tri}). We thus need to evaluate the following integral:
\begin{align} 
  I_{\rm diff} = \int_0^{\infty} dy_1 \int_{-\infty}^{y_1/\sqrt{3}} dx_1 
\int_{-\infty}^0 dy_2 \int_{-\infty}^{-y_2/\sqrt{3}} dx_2\,
  \frac{\zeta(\b r_1)\zeta(\b r_2)}{[(x_1-x_2)^2 + (y_1-y_2)^2]^3} 
\end{align}
where  
\begin{align}
  \zeta(\b r) = ( \sqrt 3 |y|-x)\frac{\tan\al}{2}
\end{align}
We first perform the $x_1, x_2, y_1$ integrals (in that order) without any cutoffs. We then evaluate the $y_2$ integral using both UV and IR cutoffs.
The small $\de/L$ expansion yields
\begin{align} \label{eq:I_diff_fin}
  I_{\rm diff} = \left(\frac{1}{128} +\frac{\pi}{72\sqrt 3}\right) \al^2 \log(L/\de) + \dotsb
\end{align}
where again the dots denote integer powers of $L/\de$ that do not affect the logarithmic trihedral contribution. We note that the sign 
of the prefactor of the logarithm is positive, in contrast to the negative one found for $I_{\rm same}$, \req{I_same_fin}. 
Using \req{diff-same-decomp}, we thus obtain the logarithmic term in the variation of the EE:
\begin{align}
  \delta S^{(2)}= \frac{C}{2} \left( \frac{9}{64} + \frac{\pi\sqrt 3}{8} \right) \al^2\log(L/\de) + \dotsb
\end{align}
which is non-negative.  
We can express the smooth limit result in terms of the angle of the faces, $\theta$:
\begin{align}
  \delta S^{(2)} = C \left( \frac{9}{32\sqrt{3} } + \frac{\pi}{4} \right) \left(\frac{2\pi}{3}-\theta \right) \log(L/\de) + \dotsb
\end{align} 
which indeed vanishes in the flat limit, $\theta=2\pi/3$.

Using the same approach, one could study trihedrals with non-equal angles, as well as $n$-hedrals where $n$ triangular
faces meet at a vertex. We leave this for future study. 

\section{R\'enyi entanglement susceptibilities}  \label{sec:Renyi}
The entanglement susceptibilities defined above for the von Neumann entanglement entropy $S$ can be generalized to the R\'enyi entropies, 
$S_n(A)=\frac{1}{1-n}\ln\tr( \rho_A^n)$. 
As before, we expand $S_n$ in terms of a small deformation of region $A$:
\begin{align} \label{eq:S-exp}
  S_n(A+\delta A)= S_n(A) + \int_{\partial A}\!\! \mbox d^{d-1}\b r\,  \chi_n^{(1)}(\b r) \zeta(\b r)
  + \frac{1}{2!}\int_{\partial A}\!\! \mbox d^{d-1}\b r \int_{\partial A}\!\! \mbox d^{d-1}\b r'\, \chi_n^{(2)}(\b r,\b r') \zeta(\b r)\zeta(\b r') 
+\cdots 
\end{align}  
The susceptibilities $\chi_n^{(\ell)}$ now depend on the R\'enyi index, $n$. Unlike for $n=1$, we can no longer assert that $\chi^{(2)}(\b r,\b r')$
is non-positive because SSA does not generalize to $n\neq 1$. However, many of the above results do generalize to $n\neq 1$ in
a straightforward way. For instance, let us focus on the case where $A$ is a half-plane in $d>1$. If we work with states $\rho$ that are invariant
under translations, rotations and scale transformations, we have   
\begin{align} \label{eq:chi_n-plane}
  \chi_n^{(2)}(\b r-\b r') = - \frac{C_n}{|\b r-\b r'|^{2d}}
\end{align}
where the constant $C_n$ reduces to $C$ as $n\to 1$, assuming the limit exists. 
In the context of CFTs, this same scaling was discussed in \cite{Faulkner15}.
$C_n$ will in general depend non-trivially on $n$
and we cannot assume that $C_n\geq 0$. \req{chi_n-plane} does imply that the results for the universal entanglement entropy of
singular surfaces derived in section \ref{sec:singular} hold for general $n$, with the replacement $C\to C_n$. 
In the nearly flat limit, we thus have the same logarithmic divergences for corners, cones and trihedral vertices, 
with universal coefficients depending on the R\'enyi index. 

In the case of CFTs, $C_n$ is related to the properties of the so-called \emph{displacement operator}, $\mathcal D(\b r)$, which generates
infinitesimal displacements of the entangling surface \cite{Bianchi16}. More precisely, 
\begin{align}
  C_n = \frac{C_{\mathcal D}(n) }{n-1}
\end{align}
where $C_{\mathcal D}(n)$ determines the 2-point function of the displacement operator, just as $C_T$ determines the 2-point function of the stress tensor.
It was further shown \cite{Bianchi16} that in $d=3$ spatial dimensions, we have    
\begin{align}
  C_n = \frac{16}{\pi^2} f_b(n) 
\end{align}
where the coefficient $f_b(n)$ appears in the EE of smooth surfaces.   

\section{Conclusion and outlook}
We have studied the entanglement susceptibilities of general quantum states. These quantify how the entanglement entropy of a spatial subregion
varies with small shape deformations of the region. In particular, we focused on the 2nd order entanglement susceptibility of a planar region
for a large family of scale invariant states. 
We used these results to derive the universal contributions that arise due to non-smooth features in the entangling surface: corners in 2d, as well as cones and trihedral vertices in 3d. In the latter case, this constitutes the first
controlled analytical result for that geometry.
Finally, we presented the generalization to R\'enyi entropies.  

First, it would be of interest to obtain the universal coefficient of the 2nd order susceptibility, $C$, for 
a larger set of critical states, such as the groundstate of fermions with a quadratic band touching or disorder driven
quantum critical points with non-trivial dynamical exponent $z$. This would shed light on the general meaning of the
universal constant $C$. 

Another rich avenue would be to study the entanglement susceptibilities for states that are not scale invariant. 
Here, an interesting starting point would be quantum Hall wavefunctions, such as Laughlin wavefunctions for both integer and fractional
fillings. Since these possess a scale, the magnetic length, the functional form of the second order susceptibility for a half-plane is no longer fixed by symmetry.

\begin{acknowledgments}     
We are very thankful to P.~Bueno, L.~Hayward Sierens, R.~Melko and R.\ C.\ Myers for many stimulating discussions.
WWK was funded by a Discovery Grant from NSERC, a Canada Research Chair, and a 
``\'Etablissement de nouveaux chercheurs et de nouvelles chercheuses universitaires'' grant from FRQNT.
Part of the work was performed at the Aspen Center for Physics, which is supported by National Science Foundation grant PHY-1066293.  
\end{acknowledgments}  

\bibliographystyle{apsrev4-1}
\bibliography{Biblio}{}    
 
\end{document}